\documentclass[12pt,oneside,peerreview,draftcls,onecolumn]{IEEEtran}
\usepackage{amsmath,amsfonts,amssymb,amsbsy, amsthm,ae,aecompl}
\usepackage{algorithm,algpseudocode}
\usepackage[utf8]{inputenc}
\usepackage[english]{babel}
\usepackage{bm}
\usepackage{color}
\usepackage[noadjust]{cite}
\usepackage{epsfig}
\usepackage{enumerate}
\usepackage{float} 
\usepackage{fancyhdr}
\usepackage[T1]{fontenc}
\usepackage[acronym,toc,shortcuts]{glossaries}
\usepackage{graphicx, caption, subcaption}
\usepackage{hyperref}
\usepackage{lastpage}
\usepackage{listings}
\usepackage{lipsum}
\usepackage{dsfont}
\usepackage{multind}
\usepackage[normalem]{ulem}
\usepackage{transparent}
\usepackage{tikz,pgfplots}
\usepackage{times}
\usepackage{verbatim}
\usepackage[all]{xy}
\usepackage{soul}

\usetikzlibrary{calc}

\hypersetup{
    colorlinks,%
    citecolor=black,%
    filecolor=black,%
    linkcolor=black,%
    urlcolor=black
}

\pgfplotsset{
  grid style = {
    dash pattern = on 0.025mm off 0.95mm on 0.025mm off 0mm, 
    line cap = round,
    black,
    line width = 0.5pt
  },
  tick label style={font=\small},
  label style={font=\small},
  legend style={font=\footnotesize},
}

\pgfplotscreateplotcyclelist{laneasBigDataMag}{
cyan!60!black,		densely dotted, 	every mark/.append style={fill=cyan!80!black},mark=otimes*\\%
red,                 solid, every mark/.append style={fill=red!80!black},mark=otimes*\\%
cyan!60!black, 		densely dotted,	every mark/.append style={fill=cyan!80!black},mark=diamond*\\%
red,           		solid, every mark/.append style={solid,fill=red!80!black},mark=diamond*\\%
teal,				every mark/.append style={fill=teal!80!black},mark=*\\%
orange,				every mark/.append style={fill=orange!80!black},mark=square*\\%
red!70!white,		mark=star\\%
yellow!60!black,		densely dashed, every mark/.append style={solid,fill=yellow!80!black},mark=square*\\%
black,				every mark/.append style={solid,fill=gray},mark=otimes*\\%
blue,				densely dashed,mark=star,every mark/.append style=solid\\%
red,					densely dashed,every mark/.append style={solid,fill=red!80!black},mark=diamond*\\%
}

\graphicspath{{figures/}}

\bgroup
\makeglossaries
\newacronym{BS}{BS}{base station}
\newacronym{CDN}{CDN}{content delivery network}
\newacronym{CELL-ID}{CELL-ID}{cell identification ID}
\newacronym{CF}{CF}{collaborative filtering}
\newacronym{CN}{CN}{core network}
\newacronym{CRP}{CRP}{{C}hinese restaurant process}
\newacronym{CS}{CS}{central scheduler}
\newacronym{D2D}{D2D}{device-to-device}
\newacronym{GGSN}{GGSN}{Gateway GPRS Support Node}
\newacronym{GPS}{GPS}{global positioning system}
\newacronym{GTP}{GTP}{GPRS Tunneling Protocol}
\newacronym{HetNet}{HetNet}{heterogeneous network}
\newacronym{HTTP}{HTTP}{Hypertext Transfer Protocol}
\newacronym{HDFS}{HDFS}{Hadoop Distributed File System}
\newacronym{HiveQL}{HiveQL}{Hive Query language}
\newacronym{ICIC}{ICIC}{inter-cell interference coordination}
\newacronym{ICN}{ICN}{information-centric network}
\newacronym{IoT}{IoT}{Internet of Things}
\newacronym{LAC}{LAC}{location area code}
\newacronym{LTE}{LTE}{long term evolution}
\newacronym{MIMO}{MIMO}{multiple-input multiple-output}
\newacronym{MO}{MO}{Mobile Operator}
\newacronym{massive-MIMO}{massive-MIMO}{massive multiple-input multiple-output}
\newacronym{MSISDN}{MSISDN}{Mobile Subscriber Integrated Services for Digital Network Number}
\newacronym{PDN}{PDN}{packet data network}
\newacronym{PoP}{PoP}{point-of-presence}
\newacronym{PPP}{PPP}{{P}oisson point process}
\newacronym{PHY}{PHY}{physical layer}
\newacronym{RMSE}{RMSE}{root-mean-square error}
\newacronym{RL}{RL}{reinforcement learning}
\newacronym{OTT}{OTT}{over-the-top}
\newacronym{SAC}{SAC}{service area code}
\newacronym{SBS}{SBS}{small base station}
\newacronym{SINR}{SINR}{signal-to-interference-plus-noise ratio}
\newacronym{SCN}{SCN}{small cell network}
\newacronym{SGSN}{SGCN}{Serving GPRS Support Node}
\newacronym{SVD}{SVD}{singular value decomposition}
\newacronym{TEID}{TEID}{tunnel endpoint identifier}
\newacronym{TL}{TL}{transfer learning}
\newacronym{UT}{UT}{user terminal}
\newacronym{URI}{URI}{Request-uniform resource identifier}
\newacronym{QoS}{QoS}{quality-of-service}
\newacronym{QoE}{QoE}{quality-of-experience}
\newacronym{RAN}{RAN}{radio access network}

\begin{document}
\title{Big Data Caching for Networking: Moving from Cloud to Edge}
\author{
		\IEEEauthorblockN{Engin Zeydan$^{\circ}$, Ejder Baştuğ$^{\diamond}$, Mehdi Bennis$^{\star}$, Manhal Abdel Kader$^{\diamond}$, Alper Karatepe$^{\circ}$, Ahmet Salih Er$^{\circ}$ and Mérouane Debbah$^{\diamond,\dagger}$				\vspace{0.4cm}} \\
		\IEEEauthorblockA{
				\small
				$^{\diamond}$Large Networks and Systems Group (LANEAS), CentraleSupélec, Université Paris-Saclay, Gif-sur-Yvette, France \\	
				$^{\star}$Centre for Wireless Communications, University of Oulu, Finland \\
				$^{\circ	}$AveaLabs, Istanbul, Turkey \\
				$^{\dagger}$Mathematical and Algorithmic Sciences Lab, Huawei France R\&D, Paris, France \\	
				\small
				\{ejder.bastug, merouane.debbah\}@centralesupelec.fr, 
				bennis@ee.oulu.fi, \\
				\{engin.zeydan, alper.karatepe, ahmetsalih.er\}@avea.com.tr, 
				manhalak@gmail.com
				\vspace{-0.4cm}
		}
		\thanks{This research has been supported by the ERC Starting Grant 305123 MORE (Advanced Mathematical Tools for Complex Network Engineering), the SHARING project under the Finland grant 128010 and TUBITAK TEYDEB 1509 project grant, numbered  9120067 and the project BESTCOM. Some technical details and procedures in this work are omitted due to space and format limitations. We refer interested readers to \cite{Bastug2015BigData} for more details.}
}
\maketitle
\begin{abstract}
In order to cope with the relentless data tsunami in $5G$ wireless networks, current approaches such as acquiring new spectrum, deploying more base stations (BSs) and increasing nodes in mobile packet core networks are becoming ineffective in terms of scalability, cost and flexibility. In this regard, context-aware $5$G networks with edge/cloud computing and exploitation of \emph{big data} analytics can yield significant gains to mobile operators. 
In this article, proactive content caching in $5$G wireless networks is investigated in which a big data-enabled architecture is proposed. In this practical architecture, vast amount of data is harnessed for content popularity estimation and strategic contents are cached at the BSs to achieve higher users' satisfaction and backhaul offloading. To validate the proposed solution, we consider a real-world case study where several hours of mobile data traffic is collected from a major telecom operator in Turkey and a big data-enabled analysis is carried out  leveraging tools from machine learning. Based on the available information and storage capacity, numerical studies show that several gains are achieved both in terms of users' satisfaction and backhaul offloading. For example, in the case of $16$ BSs with $30\%$ of content ratings and $13$ Gbyte of storage size ($78\%$ of total library size), proactive caching yields $100\%$ of users'  satisfaction and offloads $98\%$ of the backhaul.
\end{abstract}
%
\section{Introduction}
Nowadays, wireless data traffic is experiencing a tremendous growth due to pervasive mobile devices, ubiquitous social networking and resource-intensive applications of end-users with anywhere-anytime-to-anything connectivity. This unprecedented increase in data traffic  chiefly driven by mobile video, online social media and \ac{OTT}  applications are compelling mobile operators to look for innovative ways to manage their increasingly complex networks and scarce backhaul resources. In fact, a major driver of this backhaul problem is wireless video on-demand traffic in which users access contents whenever they wish in an asynchronous fashion (unlike live-streaming and digital TV), and has unique characteristics (i.e., users' demands concentrates on a small set of  popular contents, resulting in heavy-tail distribution) \cite{GSMA2015}. The explosion of data traffic stemming from diverse domains (i.e., healthcare, machine-to-machine communication, connected cars, etc.) with different characteristics (i.e., structured/non-structured) falls into the framework of \emph{big data} \cite{Lynch2008BigData}. Indeed the potential offered by big data has spurred great interest from industry, government and academics (see \cite{Hu2014TowardScalable} and references therein).

At the same time, mobile cellular networks are moving toward the next generation $5$G wireless networks, in which  ultra-dense networks, massive-\ac{MIMO}, millimeter-wave communication, edge caching, device-to-device communications are heavily investigated (see \cite{Andrews2014Will} and references therein). In contrast to the base-station centric architectures (possibly) designed for  \emph{dump} mobile terminals where requests are satisfied in a \emph{reactive} way, $5$G  networks will be user-centric, context-aware and proactive in nature.

Driven by the surge of social and mobile applications, today's mobile network architectures ought to contemplate a new paradigm shift. Indeed, the era of collecting and storing  information in data centers for data analysis and decision making has dawned. Telcos are looking for decentralized  and flexible network architectures where predictive resource management plays a crucial role, thanks to  the recent advances in storage/memory, context-awareness and edge/cloud computing \cite{paschos2016wireless, Wang2014Cache, Tao2015Content,Bonomi2012Fog}. In the wireless world, big data brings about a new kind of information sets to network planning which can be inter-connected to get a better understanding of users' behaviour and network characteristics (i.e. location, user velocity, social geo-data, etc.). 

In light of this, this work investigates the exploitation of big data in mobile cellular networks from a proactive caching point of view. Because human behaviour is highly predictable and large amount of data is streaming through operators' networks, this paper proposes a proactive caching architecture. This architecture  optimizes 5G wireless networks where large amount of available data is exploited by harnessing big data analytics and machine learning tools for content popularity estimation. We also  show how this new architecture can be exploited for caching at the edge (particularly at \glspl{BS}),  yielding higher users' satisfactions and backhaul offloading gains by moving  contents closer to users. Nowadays, it is common to have terabytes of data per second flowing in a typical mobile operator consisting of $10-20$ million subscribers, which translates into roughly exabytes monthly. As a real-world example, we process a large amount of data collected on a big data-platform from one of the major mobile networks in Turkey with $17$ millions of subscribers.  These traces are collected from several \glspl{BS} in hours of time interval and analysed inside the network to ensure privacy concerns and regulations. To the best of our knowledge, this is perhaps the first attempt to showcase the  potential of big data for caching in $5$G mobile networks.

\subsection{Prior Work and Our Contribution}
\label{sec:prior_work}
Not surprisingly, the exploitation of big data in mobile computing has been investigated recently in many works (see \cite{Shuguang2015} for example). Caching at the edge of mobile wireless networks (namely \glspl{BS} and user equipments) is also of high interest as evidenced in \cite{paschos2016wireless, FemtocachingD2D, poularakis2016caching}. Briefly, technical misconceptions of caching for $5$G networks are introduced in \cite{paschos2016wireless}. A study on improving the video transmission in cellular networks via asynchronous content reuse of cache-enabled devices is given in~\cite{FemtocachingD2D}. Cooperative caching for delivering layered videos to mobile users is studied in \cite{poularakis2016caching}.

Compared to existing works, our main contribution is to highlight and assess the potential gains of big data processing techniques for cache-enabled wireless networks, by using real traces of mobile users  collected from \glspl{BS} in a large urban area. To the best  of our knowledge, none of the previous approaches has focused on deployment of a Hadoop-based big data processing platform inside a \ac{MO}'s core network in order to validate the performance gains of caching with real-data trials. By using tools from machine learning to predict content popularity, further improvements in user's \ac{QoE} and backhaul offloading are achieved via proactive caching at the edge. 

The rest of the paper is structured as follows. The role of big data and proactive edge caching in   wireless  networks is briefly discussed in Section \ref{sec:moving}. An architecture based on big data platform and cache-enabled \glspl{BS} is proposed in Section \ref{sec:architecture}. A practical case study is carried out in Section \ref{sec:casestudy}, where the traces collected from mobile operator's network are processed on the big data platform and gains of proactive caching are validated via numerical simulations. We conclude in Section \ref{sec:concfuture} and provide future directions.
\section{Big Data Analytics for 5G Networks: Requirements, Challenges and Benefits}
\label{sec:moving}

Today's networking requirements are getting software-defined in order to be more scalable and flexible against big data.  Tomorrow's big networks will be even more  complex and interconnected. For that matter, \glspl{MO}' data centers and network infrastructures will need to monitor traffic patterns of tens of millions of clients using possible collection units of user statistics data (e.g. location, traffic demand pattern, capability, etc) for proper analysis.  

\subsection{Current Challenges and Trends in Big Data Networking}
Recently, data traffic patterns inside mobile operators' data centers have changed dramatically. Big data has enabled high traffic exchange between gateway elements at backhaul. Although wireless technology has improved tremendously from 2G to 4G, backhaul connections of cellular networks have not  seen  such a rapid  evolution. Hence, the  mobile backhaul intra-traffic is slowly becoming larger than the inter-traffic between mobile backhaul elements and end-users.  Indeed, in today's carrier networks, in addition to handling  mobile users' traffic via mobile backhaul, fetching data from a number of different backend, database and cache servers, as well as the data generated by gateway and backhaul elements also contribute to this traffic load within the operator's infrastructure. In fact, interactions of \ac{UT} triggers various interactions with hundreds of servers, routers and switches inside the backhaul and core network. For example, for an original user's HTTP request of $1$ KByte, the intranet data traffic can increase up to $930$x~\cite{Farrington2013Facebook}. This is contrary to the traditional carrier network architecture which assumes client and wireless access nodes as bottlenecks lacking computational overhead rather than the backhaul infrastructure.  Moreover, since data growth is a major challenge in today's mobile infrastructures, managing this big data-driven networks in \textit{cloud environments} is a pressing issue. For this reason,  mobile edge computing (sometimes nicknamed "Fog" computing) is yet another emerging technology where edge devices provide \textit{cloud-computing like capabilities} within the radio access network  to carry out functionalities such as communication, storage and control \cite{Bonomi2012Fog}. However for 5G networks, it should be noted that deploying distributed cloud computing capabilities near to each \glspl{BS} site (especially at locations where traffic volume is relatively low) may also increase the deployment cost considerably compared to centralised computing solutions due to availability of hundreds of sites in a typical \ac{MO}. Moreover, for modelling and prediction of spatio-temporal users' behaviour in user-centric 5G networks, network traffic arriving to a centralized location needs to be scaled out horizontally across servers and racks which is only feasible inside the core-site of a \ac{MO} for proper analysis rather than distributed locations with relatively low traffic. 
\subsection{When Big Data Analytics Meets Caching: A Hadoop Case Study}
Owing to the recent developments in networking technology and standards as well as new forms of personal communications, big data has gained increased popularity especially inside data centers and mobile operators. 
With the enormous challenges of big data inside networking world, it is evident that the only way to cope with the growing network data traffic is through better data management and movement of data from cloud into the edge. In recent years, Hadoop has been successful as a big data management software solution offering dramatic cost savings over traditional tier-one database infrastructures, processing capabilities of various data formats and parallel processing over multiple nodes. Additionally, advanced analytic techniques in machine learning in conjunction with non-relational databases that can exploit big data (e.g. NoSQL databases) have increased the opportunity of understanding big data. 

It is clear that moving contents' proximity to the edge is important whenever user's connectivity times out while performing streaming and/or downloading activities. To mitigate this, allowing data to be closer to users by reducing the distance of content to users and pushing the right content and applications at the edge yield better user experience. For instance, allowing Hadoop's distributed data processing engine for analyzing users' behaviour from enormous amount of streaming data (through the core site of \glspl{MO}) as well as exploiting proactively caching strategic contents at edges (e.g. at \glspl{BS}) can ease the backhaul traffic and improve users' \ac{QoE} by latency reduction. 
The following section discusses Hadoop-based big data processing platform and its relation with edge caching, as one way of dealing with big data inside \glspl{MO}.
\section{Big Data-Aided Cache-Enabled Architecture}
\label{sec:architecture}
\begin{figure*}[ht!]
	\centering
	\includegraphics[width=0.95\linewidth]{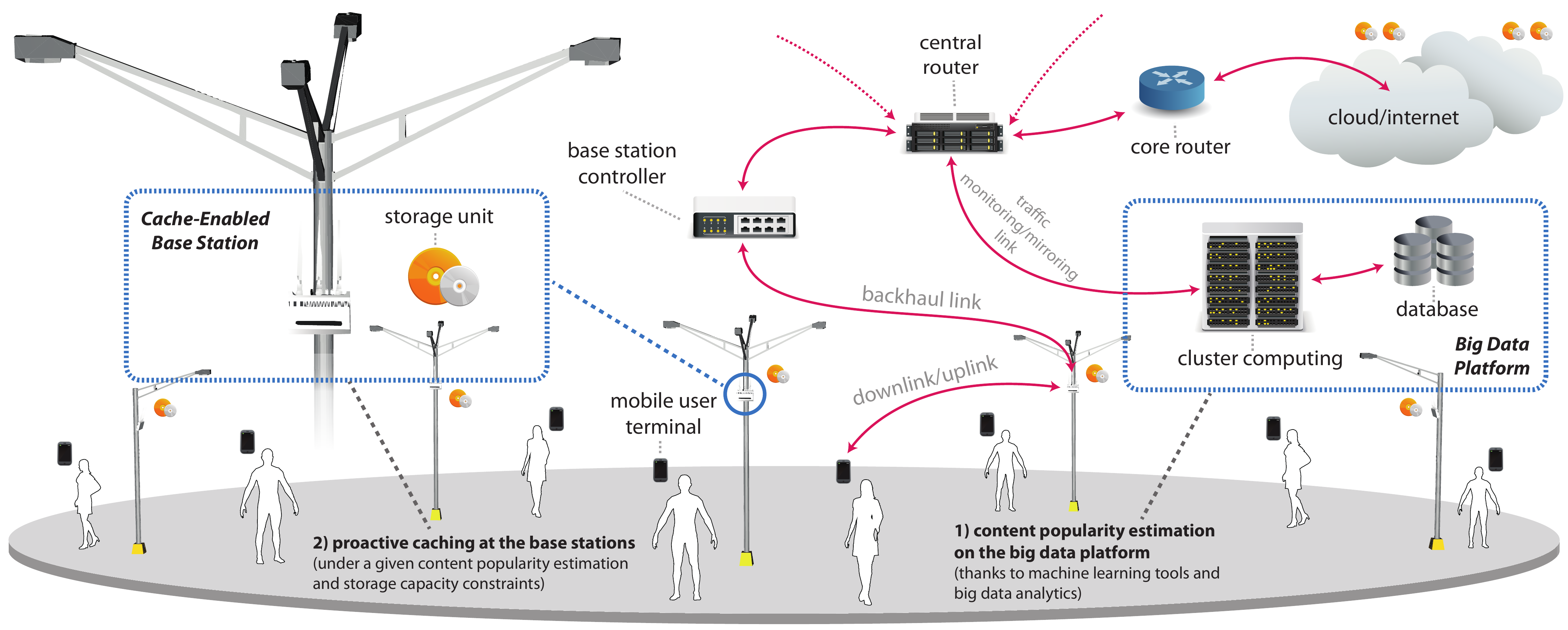}
	\caption{Illustration of the proposed architecture. The contents are moved from cloud to the edge (namely \glspl{BS}) by first inferring strategical contents on \emph{a big data platform} inside \glspl{MO} core site, then proactively storing them at the \emph{cache-enabled \glspl{BS}}.}
	\label{fig:architecture}
\end{figure*}

The  goal of this section is to investigate a new practical system architecture to gather, analyze, and proactively tackle the skyrocketing data surge. Motivated by the  highly predictable human behavior, the proposed architecture collects contextual information (e.g. user's viewing history and location information) and predicts users' spatio-temporal demand to proactively cache judiciously selected contents at the network edge. The proposed architecture parallelizes the computation and execution of the content prediction algorithms at core site and cache placement at \glspl{BS}. By doing so, users' demands are highly satisfied yielding low latency and higher \ac{QoE}. Fig. \ref{fig:architecture} shows such combined network architecture where a \emph{big data platform} deployed at core site is in charge of tracking/predicting users' demand, whereas \emph{cache-enabled \glspl{BS}} store the strategic contents predicted by the big data platform. The following sections examine the architecture details.

%
\subsection{Cache-Enabled \glspl{BS}}
\label{sec:caches}
Let us assume a small cell network  composed of $N$ small cells, where backhaul link and wireless link capacities of small cell $n$ are denoted by $C_n$ and $C'_n$ respectively. We assume that $C_n < C'_n$ reflecting  a limited backhaul capacity scenario \cite{Andrews2014Will}. A set of  users are requesting a total number of $D$ contents during $T$ time duration from a library of $\mathcal{F} = \lbrace 1, \ldots, F\rbrace$ where each content $f$ in this library has a size of $L(f)$ with $L_{\mathrm{min}} < L(f) < L_{\mathrm{max}}$ and finite bit-rate requirement of $B(f)$ during its delivery.  To offload the \textit{capacity-limited backhaul},  \glspl{SBS} are equipped with finite storage capacity and cache a subset of contents from the library $\mathcal{F}$. However, due to the sheer volume of contents and users, it is very challenging to process and extract useful information to cache users' all contents at \glspl{BS}, mainly due to  limited storage constraints and lack of sufficient backhauls.

As alluded to earlier, minimizing the backhaul load via edge caching is  very challenging. In this regard, a joint optimization of \textit{content popularity matrix} (denoted by ${\bf P}$ where columns are contents,  and rows are users or \glspl{BS} depending on the scenario)  and \textit{content cache placement} at  specific small cells  are required while considering content sizes, bit rate requirements, backhaul, etc. Moreover, limited storage capacities of \glspl{SBS}, the backhaul and wireless links, large library size and number of users with unknown ratings (i.e. empirical value of content popularity) have to be considered while dealing with a non-tractable  cache decision \cite{FemtocachingD2D}. Assuming that this non-tractable cache placement can be handled with greedy or approximate approaches (see \cite{FemtocachingD2D, poularakis2016caching}), the \glspl{SBS}  learn and estimate the sparse content popularity/ratings. The following subsection is dedicated to this task.

\subsection{Big Data Platform for Analysis}
\label{sec:analysis}
In this section, a general big data  processing framework for analyzing users' data traffic is discussed.  The purpose of this platform is to store users' data traffic  and extract useful information for proactive caching decisions. Supposing that Hadoop is deployed inside the core site of a \ac{MO}, some of the requirements of this platform for our analysis are as follows:
\begin{itemize}
\item[i)] \textbf{Huge Data Volume Processing in Less Time:}
In order to make proactive caching decisions, data processing platform inside the mobile core network infrastructure should be capable of reading and combining data from disparate data sources  and delivering intelligent insights quickly and reliably. For this reason, after mirroring the data streaming interface through network analyzing tools, the collected raw data need to be exported into a big data storage platform such as \ac{HDFS} via enterprise data integration methods (such as \textit{Spring Integration}) for detailed analysis.

\item[ii)] \textbf{Cleansing, Parsing and Formatting Data:} Data cleansing is an essential part of the data analysis process. In fact, before performing any machine learning and statistical analysis on data, data itself has to be cleaned and usually this process takes more time than the  machine learning analysis. Indeed, there are multiple steps involved in data cleansing process. First,  raw data needs to be cleaned. The raw data itself might contain some malfunctioning, inappropriate and inconsistent packets with incorrect character encodings, etc. which need to be eliminated. The next step is to extract the relevant fields from the raw data itself. In this stage, the required headers from control and data packets that will be analyzed in later stages are extracted based on the data analysis and modeling requirements. Finally, the parsed data needs to be encoded accordingly (e.g. in \textit{Avro} or \textit{Parquet} format) for appropriate storage inside \ac{HDFS}. 

\item[iii)] \textbf{Data Analysis:}  Using the formatted data in \ac{HDFS}, different data analytics  techniques can be applied over header or/and payload information of both control and data planes using high level query languages such as  \ac{HiveQL} and Pig Latin. The aim of such a step is to find  relationships between control and data packets, e.g. the location or \ac{MSISDN} of users (that are present in control packets but not in data packets) to the requested content (that are present only in data packets) through successive Map-Reduce operations. 

\item[iv)] \textbf{Statistical analysis and visualizations:} After machine learning analysis is done to predict the spatio-temporal user behaviour for proactive caching decisions,  the results of the analysis can be stored and reused. Moreover, the results can be re-formatted to be used for further analysis using appropriate Extract, Transform and Load (ETL) tools, and  can be input to other processing systems such as Apache Spark's MLlib, etc. In addition, visualizations such as graphs and tables can be used to represent the data in a visual format for ease of understanding. 
\end{itemize}

In such a platform, machine learning techniques which lie at the heart of recommendation-based engines can be applied so that users' strategic contents can be inferred from a large amount of available data. 
In what follows, as a practical case study, we first analyse huge amount of users' traffic on such a big data platform and use this data to estimate the content popularity matrix ${\bf P}$, which is essentially required for caching decision. Subsequently, we conduct a numerical study for showcasing the gains of caching at \glspl{BS}.
%
\section{A Practical Case Study}
\label{sec:casestudy}
Data streaming traces for this practical case study are collected from one of the regional core district of mobile operator's network which consists of more than $10$ regional core areas in Turkey. A mirroring procedure is initialized for transferring streaming traces into the big data platform in the core network. A fast speed of $200$ Mbit/sec at peak hours is observed through one of the mirrored interfaces in the core network. The total average traffic over all regional areas consists of approximately over $15$ billions of packets in the uplink and over $20$ billions of packets in the downlink direction daily. This is equivalent to almost $80$~TByte of total data.\footnote{In fact, the general trend of data in the network is following an exponential growth, i.e., in $2012$, the total average data traffic per day was over $7$ TByte in both uplink and downlink.} This mirrored network traffic is analyzed on the data processing platform which is essentially based on Hadoop. In particular, the big data platform is composed of Cloudera's Distribution Including Apache Hadoop (CDH4) version on four nodes including one cluster name node, with each node empowered with INTEL Xeon CPU E5-2670 running @2.6 GHz, $32$ Core CPU, $132$ GByte RAM, $20$ TByte hard disk. As stated before, the platform is in charge of extracting the useful information from raw data. In our analysis, the traffic of approximately $7$ hours (starting from $12$ pm to $7$ pm on Saturday $21$'st of March $2015$) is collected.\footnote{Note that the size of raw data is around $1.2$ TByte for observed time duration $T$, and for offline processing, it can take up-to five days to extract the relevant headers from this data using a single server.} The traces processed on the big data platform have approximately four millions of HTTP content requests, and stored in a comma-separated text file format after following steps (i) and (ii) as described  in Section \ref{sec:analysis}. After some post processing (i.e., calculating content sizes), the \emph{final-traces} table/file which includes arrival time (abbreviated as FRAME-TIME), requested content (abbreviated as HTTP-URI) and content size (abbreviated as SIZE) is obtained, and is used in the rest of this study. The detailed description of the data extraction process is given in \cite{Bastug2015BigData}. Note that the data extraction process is specific to our scenario for proactive caching. However, similar studies in terms of usage of big data platform and exploitation of big data analytics for telecom operators can be found in the literature (see \cite{Celebi2013BigData} for instance).
%
\subsection{System Parameters and Studied Methods}
%
\begin{table}[ht]
\centering
\scriptsize
\caption{List of simulation parameters.\label{tab:setup_params1}}
\begin{tabular}{|c|l|c|}
\hline
\textbf{Parameter}			  &  \textbf{Description} 	& 	\textbf{Value}\\
\hline
$T$    		& Time duration							& 	$ 6$ hours $47$ minutes \\
\hline
$D$ 			& Nr. of requests 				& 	$422529$ 	\\
\hline
$F$				& Nr. of contents 			&  $16419$	 \\
\hline
$N$   			& Nr. of small cells 				& 	$16$ \\
\hline
$L_{\mathrm{min}}$  			& Min. size of a content & 	$1$	Byte \\
$L_{\mathrm{max}}$  			& Max. size of a content & 	$6.024$	GByte \\
\hline
$B(f)$ 			& Bitrate of content $f$				&	$4$ Mbyte/s	\\
\hline
$\sum_{n}{C_n}$ 			& Total backhaul link capacity		&	$3.8$	Mbyte/s	\\
\hline
$\sum_{n}C'_{n}$  	& Total wireless link capacity 		&	$120$	 Mbyte/s\\
\hline
\end{tabular}
\end{table}

In the numerical setup,  we assume that $D$ contents are requested from the processed data (namely from \emph{final-traces} table) over a time interval of $6$ hours $47$ minutes. Information on FRAME-TIME, HTTP-URI and SIZE are also taken from the \emph{final-traces} table. Then, the requests are pseudo-randomly assigned to $N$ \glspl{BS}. The wireless link capacities of small cells, backhaul link and storage capacities are set to identical values within each other for ease of revealing the caching gains. The list of simulation parameters are summarized in Table \ref{tab:setup_params1}. The global procedure contains the following two major steps:
	\begin{itemize}
		\item \emph{Estimation of content popularity ${\bf P}$ (where columns are contents,  and rows are \glspl{BS}):} This is done on the big data platform by processing large amount of collected data and exploiting machine learning tools. Two methods are examined in the numerical setup:
		\begin{itemize}
	        \item[i)] \textbf{\emph{Ground Truth}}: The ${\bf P}$ matrix is constructed by considering all available information in the \emph{final-traces} table.  The matrix has $6.42\%$ of rating density in total.
	        \item[ii)] \textbf{\emph{Collaborative Filtering}}: $30\%$ ratings available in the \emph{final-traces} table are picked uniformly at random for training of ${\bf P}$ matrix estimation. Then, the remaining missing entries/ratings in the traces are predicted via the regularized \ac{SVD} from \ac{CF} methods \cite{Lee2012CF}.\footnote{Regularized \ac{SVD} is chosen due to its outperforming performance with respect to other \ac{CF} methods~\cite{Lee2012CF}.}
        \end{itemize}
		\item \emph{Caching strategic contents:} The cache decision procedure at the base stations is made by storing the most-popular contents greedily at the \glspl{SBS} until no storage space remains as in~\cite{Bastug2015BigData}.\footnote{This greedy approach is chosen for ease of exposition. Once can also employ online cache placements strategies (e.g. least recently used (LRU), least frequently used (LFU)).}.
	\end{itemize}

As regards to the performance metrics: i) \emph{request satisfaction}, as \ac{QoE} metric, is defined as the amount of contents delivered at a given target rate, and ii) \emph{backhaul load} corresponds to the percentage of the traffic passing over the backhaul links over the total possible traffic volume induced by the content requests. A detailed analytical formulas of \emph{request satisfaction} and \emph{backhaul load} can be found in ~\cite{Bastug2015BigData}.

\subsection{Numerical Results and Discussions}
\begin{figure*}[!ht]
\centering
\begin{subfigure}{.48\textwidth}
\begin{tikzpicture}
	\begin{axis}[
		width=\textwidth,
    every tick label/.append style  =
    { 
        font=\scriptsize
    },
 		grid = major,
		legend columns=2,
		legend entries={Ground Truth, Collaborative Filtering},
		legend cell align=left,
		legend style ={font=\scriptsize},
		legend to name=namedmethods4telco,
 		mark repeat={2},
 		xmin=0,xmax=100,	
 		xlabel={Storage Size (\%)},
 		ylabel={Satisfaction (\%)}]
 		\addplot+[smooth, blue!60!black, mark=o, mark size=1.8, thick]
 				  table [col sep=comma] {\string"results-storage-satisfaction-c1ground.csv"};
 		\addlegendentry{Ground Truth};

		\addplot+[red!80!black, dashed, very thick, mark=none] 
                                 table [col sep=comma] {\string"results-storage-satisfaction-c2cf.csv"};
		\addlegendentry{Collaborative Filtering};
		 		  		
	\end{axis}
\end{tikzpicture}
\end{subfigure}
\begin{subfigure}{.48\textwidth}
\begin{tikzpicture}
	\begin{axis}[
		width=\textwidth,
    every tick label/.append style  =
    { 
        font=\scriptsize
    },
 		grid = major,
 		mark repeat={2},
 		xmin=0,xmax=100,	
 		xlabel={Storage Size (\%)},
 		ylabel={Backhaul Load (\%)}]

 		\addplot+[blue!60!black, mark=o, mark size=1.8, thick]
 				  table [col sep=comma] {\string"results-storage-backhaul-c1ground.csv"};

		\addplot+[red!80!black, dashed, very thick, mark=none] 
                                 table [col sep=comma] {\string"results-storage-backhaul-c2cf.csv"};
		 		  		
	\end{axis}
\end{tikzpicture}
\end{subfigure}
\vspace{0.2cm}
\ref{namedmethods4telco}
\caption{Numerical results for proactive caching at the base stations.}
\label{fig:satback}
\vspace{-0.6cm}
\end{figure*}
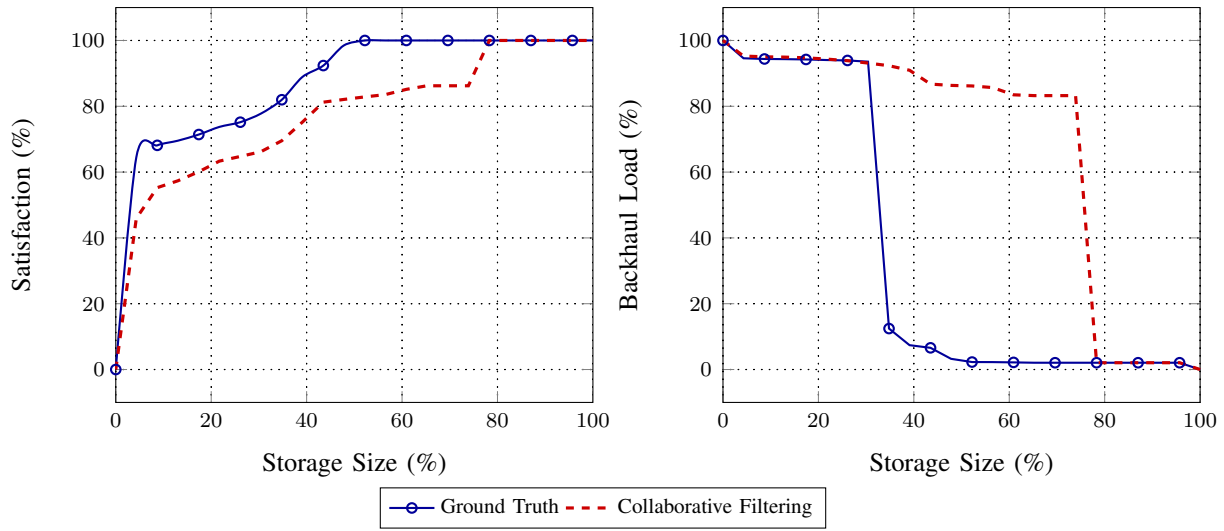

In this section, based on the available information in the \emph{final-traces} table, we conduct a numerical study to reveal the gains of caching. The impact of storage size on the users' request satisfaction is plotted in Fig.~\ref{fig:satback}. Therein, $0\%$ of storage size corresponds to no caching, whereas $100\%$ of storage is equivalent to caching the entire library ($17.7$ GByte). In the figure, we note that the users' request satisfaction has a monotonically increasing behaviour, and somewhat intuitive, $100\%$ of satisfaction is achieved in both methods when the complete content catalog (with $100\%$ of storage size) is stored by fixing parameters in our setting to plausible (and realistic) values in order to see the regimes where $100\%$ satisfaction is achieved. However, a performance gap between the ground truth and \ac{CF} is observed until $79\%$ of storage size which is mainly due to the estimation errors. For instance, when the \glspl{BS} have $40\%$ of storage size for caching, the ground truth yields $89\%$ of satisfaction whereas the performance of \ac{CF} stays at $75\%$.

Fig. \ref{fig:satback}  also shows the impact of storage size on the backhaul load/usage. In the figure, we see that both methods yield less backhaul load (namely higher offloading gains). For instance, having $79\%$ of storage size at \glspl{BS}, both methods offload $98\%$ of backhaul. However, the ground truth outperforms the \ac{CF} method since it has complete information of the content ratings. On the other hand, after a certain storage size, a dramatical decrease of backhaul is observed in both approaches. Compared to previous works which mostly consider identical content sizes, we are dealing with real traces with non-identical content sizes. 

In the simplest form, one can write the backhaul load of a particular content as $load = popularity \times size$, if not cached. Therefore, a relatively less popular but very big-sized contents might lead to such a behaviour on the backhaul load, if not cached at the \glspl{SBS}. This points out the importance of taking into account contents sizes in caching decision which reflects a more practical/realistic characterization of backhaul usage.

\begin{figure}[!ht]
\centering
\begin{tikzpicture}
	\begin{axis}[
		width=0.50\columnwidth,
    every tick label/.append style  =
    { 
        font=\scriptsize
    },
 		grid = major,
 		xlabel={Normalized backhaul Capacity (\%)},
 		ylabel={Satisfaction (\%)}]

 		\addplot+[blue!60!black, mark=o, mark size=1.8, thick]
 				  table [col sep=comma] {\string"results-imbalance-satisfaction-c1ground.csv"};

		\addplot+[red!80!black, dashed, very thick, mark=none] 
                                 table [col sep=comma] {\string"results-imbalance-satisfaction-c2cf.csv"};
		 		  		
	\end{axis}
\end{tikzpicture}
\caption{Evolution of satisfaction with respect to the normalized backhaul capacity.}
\label{fig:link}
\end{figure}
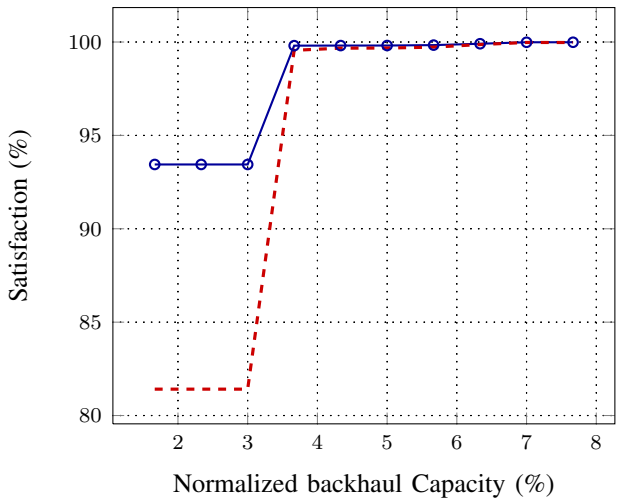

Fig. \ref{fig:link}  illustrates the evolution of users' request satisfaction with respect to the backhaul capacity ratio, defined as the ratio of total backhaul link capacity $\sum_{n}{C_n}$ over total wireless link capacity $\sum_{n}{C'_n}$. It is clear from the figure that increasing the backhaul link capacity yields higher satisfactions, both in ground truth and \ac{CF} approaches. This is due to the fact that bottleneck in the backhaul becomes less relevant with the increment of this ratio.

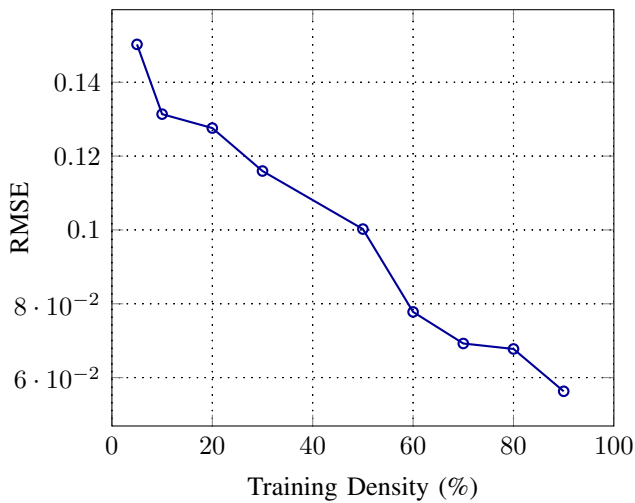
\begin{figure}[!ht]
\centering
\begin{tikzpicture}
	\begin{axis}[
		width=0.50\columnwidth,
 		grid = major,
 		legend cell align=left,
 		xmin=0,xmax=100,	
 		legend style ={legend pos=north east},
 		xlabel={Training Density (\%)},
 		ylabel={RMSE}]

 		\addplot+[blue!60!black, mark=o, mark size=1.8, thick]
 				  table [col sep=comma] {\string"results-rmse-satisfactionRatio.csv"};
		 		  		
	\end{axis}
\end{tikzpicture}
\caption{Change of the \ac{RMSE} with respect to the training density.}
\label{fig:rmse}
\end{figure}

The above performance results demonstrate the case with $30\%$ of rating density in \ac{CF}. However, it is clear that increasing the training rating density of \ac{CF}, less estimation error and hence closer satisfaction gains to ground truth is expected. In order to show this, Fig. \ref{fig:rmse} demonstrates the effect of training rating density on \ac{RMSE} where the error is defined as the root-mean-square of the difference between users' content satisfaction of the ground truth and \ac{CF} approaches over all possible storage sizes. Fig. \ref{fig:rmse} clearly validates the fact that performance of \ac{CF} can be improved via higher training rating density.
\section{Conclusions and Future Work}
\label{sec:concfuture}
In this paper, we have introduced a proactive caching architecture for $5$G wireless networks by processing a huge amount of available data on a big data platform, and leveraging machine learning tools for content popularity predictions. Additionally, relying on this prediction and using extracted traffic information from this data, the gains of caching have been investigated throughout numerical studies. One possible direction of this work is to investigate the proposed big data analysis framework in a real-time fashion. For this, recent frameworks that exist in Hadoop eco-system such as Apache Spark and its built-in libraries Spark Streaming for real-time data processing and MLLib for machine learning libraries are of interest. 
\bibliographystyle{IEEEtran}
\bibliography{references}
\end{document}